# Effect of sample size on magnetic $J_c$ for MgB$_2$ superconductor

J. Horvat,[a)] S. Soltanian, X. L. Wang and S. X. Dou,

Institute for Superconducting and Electronic Materials, University of Wollongong, NSW 2522, Australia

Abstract

A strong effect of sample size on magnetic $J_c(H)$ was observed for bulk MgB$_2$ when $J_c$ is obtained directly from the critical state model. Thus obtained zero-field $J_c$ ($J_{c0}$) decreases strongly with the sample size, attaining a constant value for the samples larger than a few millimetres. On the other hand, the irreversibility field ($H_{irr}$) defined at $J_c$ = 100 A/cm$^2$ increases with the sample size. The decrease of $J_{c0}$ is described in terms of voids in the bulk MgB$_2$ samples and superconducting screening around the cells of superconducting material between these voids (35 µm), because of concentration of the current in the narrow bridges connecting the cells. For samples larger than a few millimetres, the value of magnetic $J_c$ is in agreement with the transport $J_c$ and it is restricted by the voids. The critical state model is not suitable for obtaining $J_c$ for small bulk MgB$_2$. The increase of $H_{irr}$ with the sample size is an artefact of defining $H_{irr}$ by the value of $J_c$ at which an additional superconducting screening on 1µm scale dominates $\Delta m$.

a) e-mail: jhorvat@uow.edu.au

Superconducting wires based on $MgB_2$ are currently in the process of fast development, resulting in improved values of $J_c$, $J_c(H)$ and $H_{irr}$ [1-6]. We will show that the common practice of obtaining $J_c$ from the critical state model gives erroneous values of $J_c$ without proper analysis of superconducting screening at different length-scales.

The field dependence of $J_c$ was obtained from measurements of magnetic hysteresis loops, using a critical state model for appropriate geometry and the dimensions of the whole sample. The value of $H_{irr}$ was obtained as the field at which thus obtained $J_c$ equals to 100 A/cm$^2$. We will show that thus obtained values of $J_c$ and $H_{irr}$ are incorrect, but will use them to demonstrate our argument. The measurements were performed by a Quantum Design PPMS magnetometer, with the sweep rate of the field of 50 Oe/s. Two groups of samples were measured. An $MgB_2$ pellet was prepared by reacting magnesium and boron powders at 850°C under isostatic pressure of 150 MPa for 1 hour. The density of thus prepared pellet was 1.9 g/cm$^3$. The $T_c$ of 38.9K was obtained from measurements of ac susceptibility, with the transition width lower than 1K. The pellet was cut into a rectangular rod and measured. Subsequent measurements were performed after reducing all three dimensions of the same sample by 20%. In this way, any geometrical effects on our results were eliminated. The sample dimensions are shown in Table 1. The field was applied along the longest dimension of the sample, $x$. The second group of samples were round iron sheathed $MgB_2$ wires, prepared by powder in tube method, described elsewhere [7]. The iron sheath was removed before the measurements. Two groups of the wires were measured (Table 1). In the group D, the length of the wire was kept constant and its diameter was decreased for each of the measurements. In the group Z, the diameter was kept constant and its length was decreased. For each of the measurements, the field was applied along the cylindrical $z$-axis of the wire.

Figure 1 shows the field dependence of $J_c$ at 5 and 20K, for a series of samples with subsequently decreasing volume. The $J_c$ for low fields at 5K could not be calculated because of the flux jumps [8]. There was a strong influence of the sample size on $J_c(H)$ as well as on the zero-field $J_c$ ($J_{c0}$). For example, the $J_c$ dropped by more than two orders of magnitude at 7 T and 5K when the sample dimensions decreased from 25 to 0.26mm$^3$. This was accompanied by an increase of $J_{c0}$ as the sample size decreased.

The increase of $H_{irr}$ with the volume ($V$) of the sample is shown in Figure 2 for T = 20K. $H_{irr}$ increased very fast with $V$ for $V < 3.5$mm$^3$, followed by a much more gradual increase for $V > 3.5$mm$^3$. The transition between these two regimes occurred at the same volume for all temperatures, however $H_{irr}$ vs. $V$ for different temperatures were not scaleable to a unique curve. Inset to Figure 2 shows the field dependence of $J_c$ for two different MgB$_2$ samples at 5K, obtained from magnetic and transport measurements. The voltage contacts in transport measurements were at a distance of 1 cm, whereas the sample size for magnetic measurements was 3.5 x 3 x 0.5 mm$^3$. There is a good agreement between the two types of measurements.

Figure 3 shows the dependence of $J_{c0}$ on the sample volume. Inset to Figure 3 shows the field dependence of $J_c$, detailing the increase of $J_{c0}$ as the sample volume was decreased. Normalising $J_{c0}$ to its value at $V = 12.87$ mm$^3$, the experimental points obtained at 10, 20 and 30 K overlapped. As the sample volume decreased from 25 to 6.9 mm$^3$, the value of $J_{c0}$ increased by only 10%. However, the increase of about 60 % of the initial $J_{c0}$ was obtained as the volume decreased from 6.9 to 0.25 mm$^3$. This implies that the reported high values of $J_{c0}$ for samples with V < 1mm$^3$ should not be directly compared with $J_{c0}$ obtained for larger samples.

Figure 4 shows the dependence of $J_{c0}$ on the length of the wire (samples Z), therefore on the dimension parallel to magnetic field. $J_{c0}$ again decreases with the length of the sample. Inset to Figure 4 shows the dependence of $J_{c0}$ on the diameter of the round wire (samples D), with the field along its z-axis. $J_{c0}$ also decreases with the sample diameter. This excludes the change of the electrical field (*E*) as the origin for the sample size dependence of $J_{c0}$. Namely, for a cylindrical sample with magnetic field along its z-axis: $E = -D/2*dB/dt$ on its surface, where *D* is the diameter of the sample. However, our measurements show that $J_{c0}$ also changes with the length of the cylinder (Fig.4), where *E* remains constant.

The explanation for the sample size dependence of $J_{c0}$ should instead be sought in the terms of sample homogeneity. The density of bulk $MgB_2$ is about 70% of the density of $MgB_2$ crystal. This is mainly due to the voids scattered through the superconducting matrix [9]. The volume occupied by the superconductor can be divided into bulkier cells and narrow bridges connecting the cells. The cells are formed by the superconducting material in the volume where the voids are the furthest away from each other (Fig.5). The bridges are formed where the neighbouring voids are the closest to each other [9]. The average size of the cells is ~ 35 μm. The cells and bridges themselves consist of well-connected crystals with two tiers of agglomeration [10]. Of interest to us are the agglomerates of crystals of the order of 1 μm, whereas the agglomerates on 10 nm scale can be neglected.

The cells and bridges consist of the same material and have the same $J_{c0}$. However, the screening currents flowing around the sample are concentrated in the bridges, resulting in a smaller current density in the cells than in the bridges (Fig.5). Because the maximum current density in the bridges is $J_{c0}$, additional screening of the cells is possible, making the net current density in the cells also equal to $J_{c0}$. This resembles the inter- and intra- grain currents in high-temperature superconductors, with a significant difference that the bridges between the cells in high quality $MgB_2$ are not Josephson junctions. The sample size dependence of magnetically obtained $J_c$ can be explained by a model devised for high-temperature superconductors [11]. Approximating the shape of the sample and cells by round cylinders, the irreversible magnetic moment obtained from the hysteresis loop is [11]:

$$\Delta m = (afJ_{cc} + DJ_{cs})2V/3, \qquad (1)$$

where $J_{cc}$ and $J_{cs}$ are the current densities of the small loops around the cells and of the overall sample screening, respectively. $D$ is the diameter of the sample, whereas $a$ and $f$ are the typical diameter of the cells and their filling factor, respectively. Because $J_{cs}$ is restricted by the current concentrating in the bridges, $J_{cc} > J_{cs}$.[12] The contribution of $J_{cc}$ to $\Delta m$ is negligible for large samples ($D >> af$ in Eq. (1)), resulting in a sample size independent $J_c$ equal to $J_{cs}$. With lowering the sample size, the contribution of $J_{cc}$ starts increasing. Further, the length scale used for calculating $J_c$ in the critical state model (i.e. $D$) starts approaching the size of the cells. Because of this, and because $J_{cc} > J_{cs}$, thus calculated value of $J_c$ increases with decreasing the sample size. As the sample approaches the size of the cells, $J_{cs}$ decreases because the bridges between the cells are broken up. Finally, for $D = a$, the calculated $J_c$ equals $J_{cc}$. The simple critical state model cannot be used to calculate $J_c$ when both $J_{cc}$ and $J_{cs}$ contribute significantly to $\Delta m$. The solid line in Fig. 3 shows the fit of normalised $J_{c0}$ vs. $V$, using $J_{c0} = 3\Delta m/(2VD)$ and $\Delta m$ from Eq. (1). From here, $J_{c0} = J_{cc}af/D + J_{cs}$, where $D \approx V^{1/3}$. Microscopic examination of the samples shows that the average size of the cells is about 35 µm and the density of the sample gives $f = 0.7$, which results in $af = 25$ µm. The value of $J_{cc}/J_{cs} = 30$ was obtained by the fitting [12], giving[13] $J_{cc} \approx 5 \times 10^6$ A/cm$^2$. The dependence of $J_{c0}$ on sample length is also explained by the presence of voids. Namely, the current meanders between them following the bridges not only perpendicular to the field, but also parallel to it. Decreasing the sample length terminates the current paths that incorporate the bridges parallel to the field.

Figures 2 and 3 show that $J_{c0}$ and $H_{irr}$ have the opposite dependence on the sample volume, which is seemingly contradicting. However, Figure 1 shows that the values of $H_{irr}$ were obtained from the part of the $J_c(H)$ curve exhibiting an inflection and a step. This step in $J_c(H)$ was explained by superconducting screening around agglomerates of crystals of size ~200 μm [8]. TEM examination revealed that the size of these agglomerates is ~ 0.2 - 1 μm for higher quality samples, as the ones measured here [10]. Our measurements show that the length-scale of superconducting screening at $H > H_{irr}$ for the samples reported here is ~ 1 μm [13]. This indicates that the sample size dependence of $H_{irr}$ in Fig. 2 is an artefact of the transition to the dominant contribution into $\Delta m$ of this screening around ~1 μm large agglomerates of crystals inside the cells for $H > H_{irr}$. The mechanism is the same as for the screening around ~ 35 μm large cells, that leads to the size dependence of apparent $J_{c0}$ and to an inflection in $J_c(H)$ at about 2 T (Fig.1) [13].

Acknowledgments: Authors thank D. Larbalestier, T. Collings and E. H. Brandt for discussion and Australian Research Council, Hyper Tech Research Inc. and Alphatech International for financial support.

Figure Captions

Figure 1: Field dependence of $J_c$ for samples of different size (Table 1).

Figure 2: Dependence of the irreversibility field on the sample volume for the rectangular samples a-h at $T = 20$K. Inset: Field dependence of $J_c$ for $MgB_2$ samples at 5K, obtained from magnetic (solid line) and transport (open symbols) measurements.

Figure 3: Dependence of the normalised $J_{c0}$ on the sample volume for samples a-h (Table 1). $J_{c0}$ for $T = 10$, 20 and 30 K was normalised to its value for $V = 12.87$ mm$^3$. Solid line is fit with Eq.(1), using $D \approx V^{1/3}$: $J_{c0} = J_{cc}afV^{1/3} + J_{cs}$. Inset: Field dependence of $J_c$ showing the increase of $J_{c0}$ as the volume decreases.

Figure 4: Dependence of $J_{c0}$ on the sample length of the $MgB_2$ wire, where the diameter of the wire did not change (samples Z in Table 1). Inset: Dependence of $J_{c0}$ on the diameter of the $MgB_2$ wire, with fixed wire length (samples D in Table 1).

Figure 5: Schematic drawing of the screening currents in the sample. $J_{cs}$ and $J_{cc}$ are drawn by solid and dotted lines, respectively. The screening currents on 1 μm length-scale are represented by the dots. The shaded ellipses are the voids in the sample.

Table 1: The dimensions of the samples measured. Samples a-h are rectangular pellets whose all three dimensions (*x*, *y*, *z*) were reduces by 20% before each subsequent measurement. The field was oriented along *x*. Samples Z are round wires whose length was reduced before each subsequent measurement. The field was along the length of the samples, i.e. along the cylindrical *z*-axis. Samples D were round wires with constant length, but the diameter was reduced before each subsequent measurement. The field was also along the cylindrical *z*-axis.

| sample | x (mm) | y (mm) | z (mm) | V (mm$^3$) |
|---|---|---|---|---|
| a | 7.15 | 3.27 | 1.07 | 25.01 |
| b | 5.72 | 2.62 | 0.86 | 12.87 |
| c | 4.65 | 2.12 | 0.7 | 6.90 |
| d | 3.64 | 1.68 | 0.57 | 3.49 |
| e | 2.92 | 1.34 | 0.46 | 1.78 |
| f | 2.29 | 1.08 | 0.36 | 0.89 |
| g | 1.87 | 0.85 | 0.29 | 0.46 |
| h | 1.42 | 0.68 | 0.24 | 0.23 |

| sample | Diameter (mm) | Length (mm) |
|---|---|---|
| D1 | 1.54 | 3.91 |
| D2 | 1.23 | 3.91 |
| D3 | 0.93 | 3.91 |
| Z1 | 1.54 | 6.23 |
| Z2 | 1.54 | 3.91 |
| Z3 | 1.54 | 1.95 |
| Z4 | 1.54 | 1.50 |

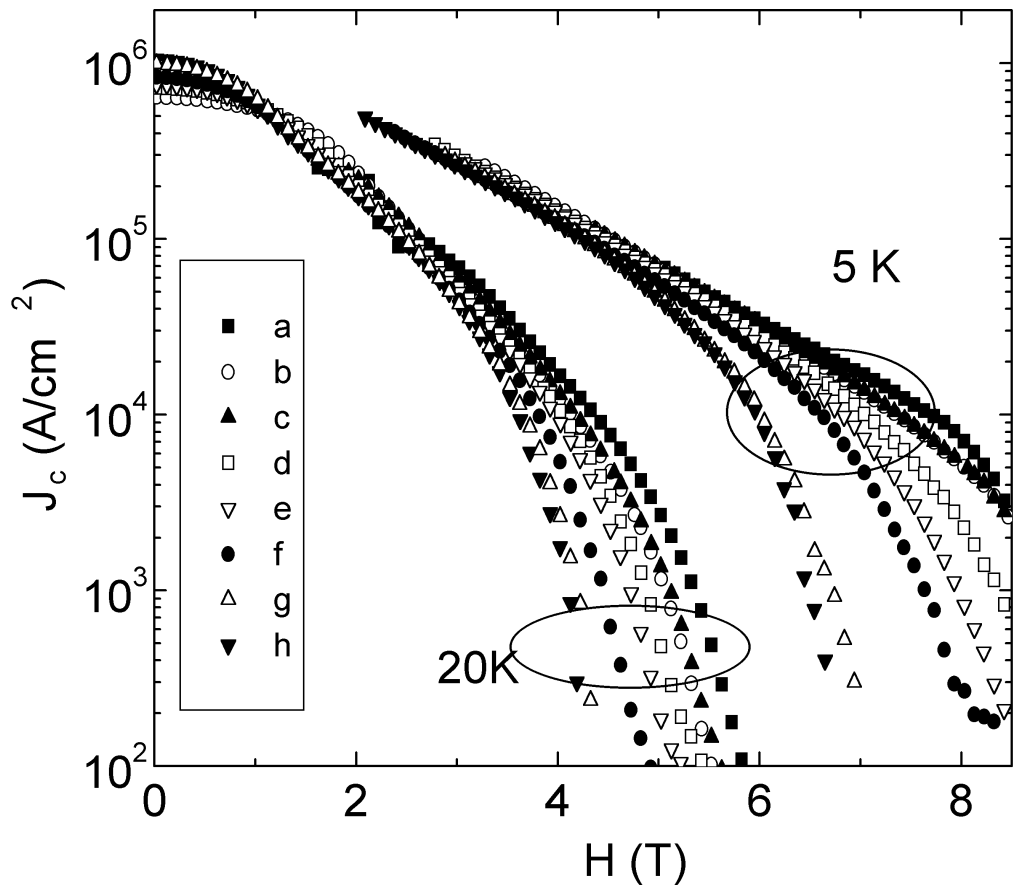

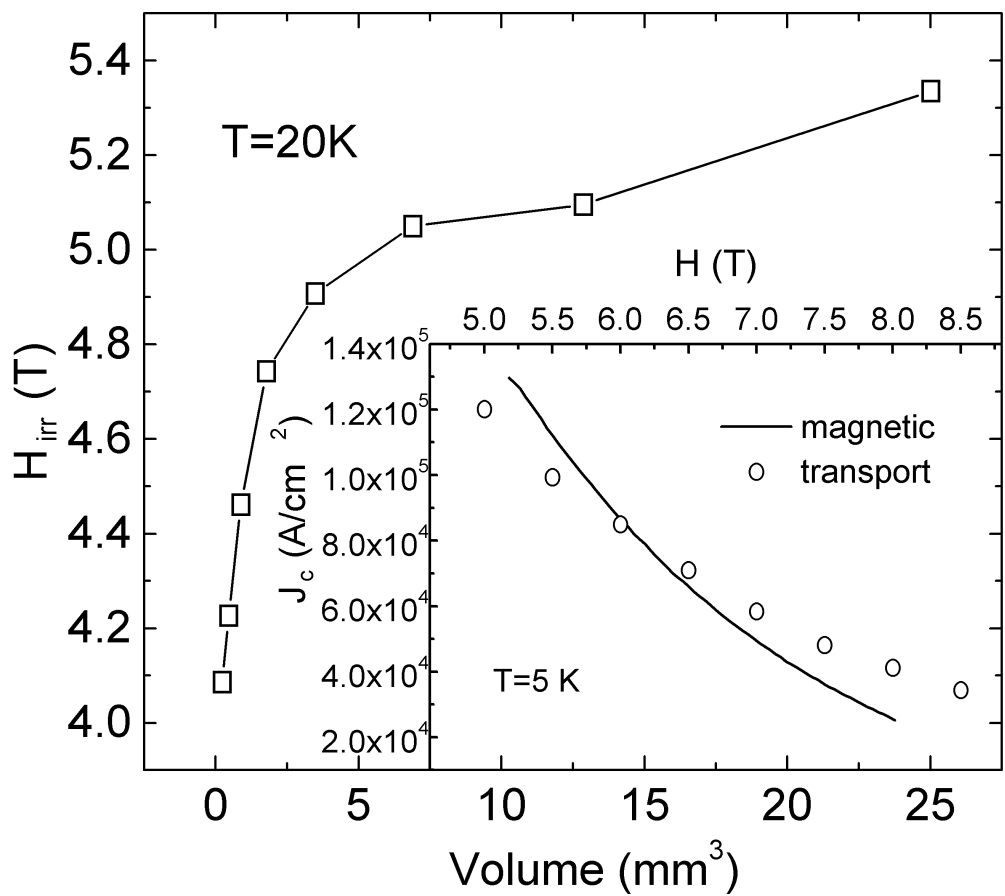

Figure 3: J. Horvat et al.

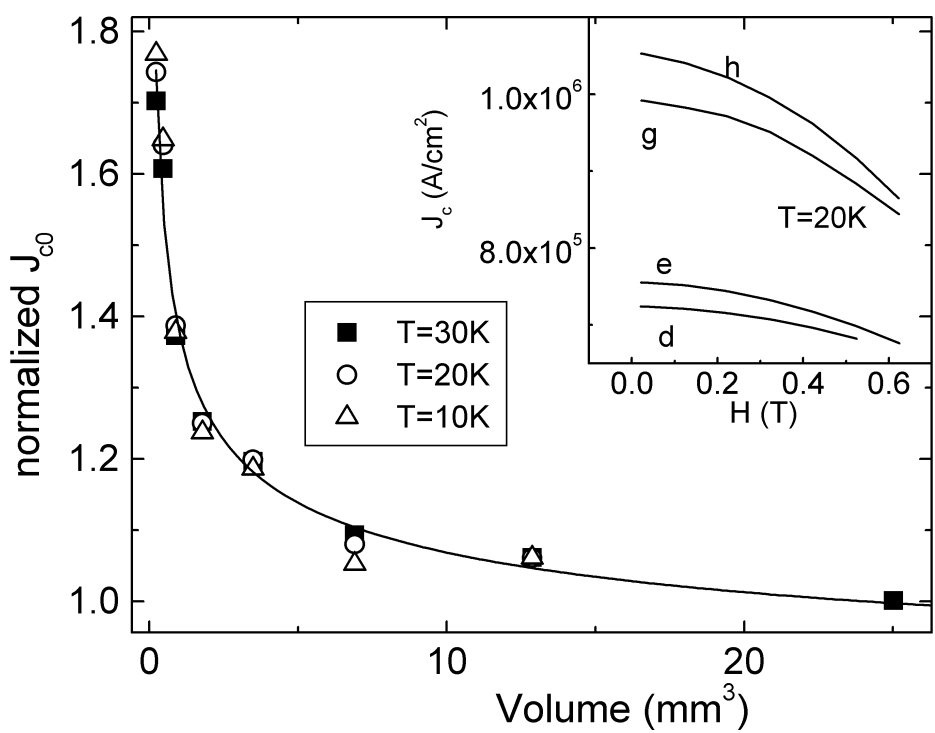

Figure 4: J. Horvat et al.

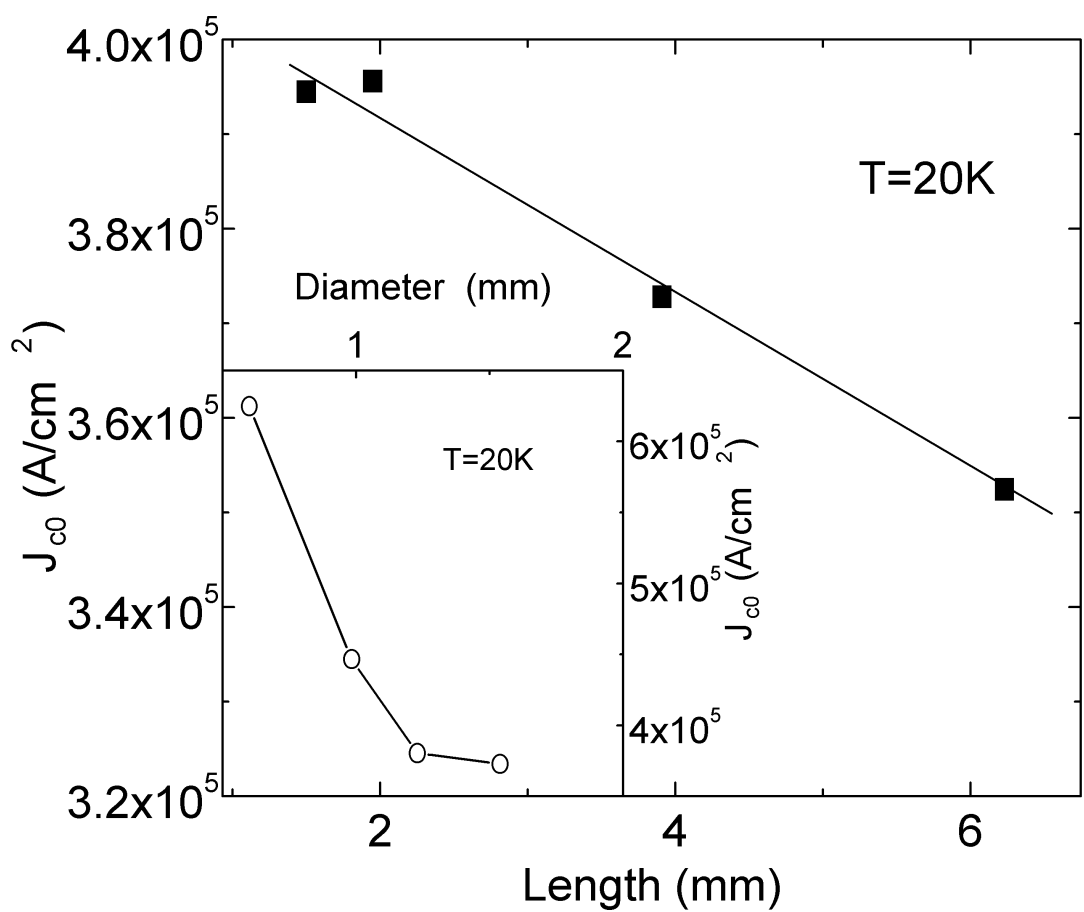

Figure 5: J. Horvat et al.

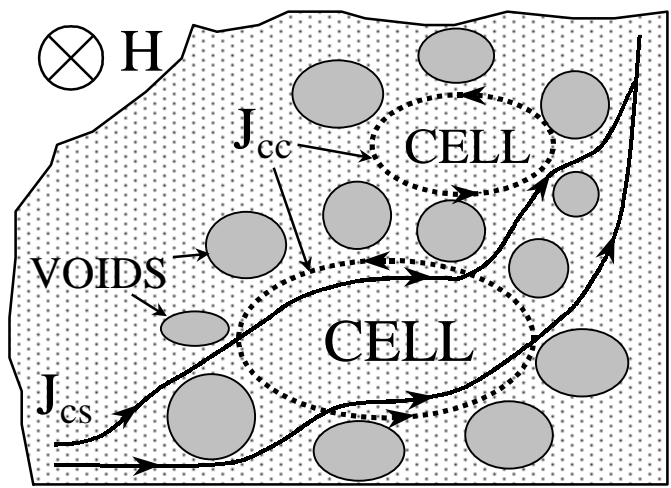